\documentstyle[preprint,prl,epsf,eqsecnum,aps,floats,amssymb]{revtex}
\input epsf

\newcommand{\pbarp}{\mbox{$\bar{p}p$}}
\newcommand{\ipb}{\mbox{${\rm pb}^{-1}$}}
\newcommand{\jjmass}{\mbox{$M$}}
\newcommand{\Et}{\mbox{$E_{T}$}}
\newcommand{\modeta}{\mbox{$\mid \! \eta  \! \mid$}}
\newcommand{\modetajet}{\mbox{$\mid \! \eta_{\rm jet}  \! \mid$}}

\newcommand{\Etu}[1]{\mbox{$E_{T}^{{\rm jet}#1}$}}
\newcommand{\Etmax}{\mbox{$E_{T}^{\rm max}$}}
\def\gevcc{GeV/$c^2$}                   
\newcommand{\chisq}{\mbox{$\chi^{2}$}}
\newcommand{\HT}{\mbox{${S_{T}}$}}

\begin{document}
\tightenlines
\preprint{Fermilab-Pub-98/220-E,hep-ex/9807014}
\title{
The Dijet Mass Spectrum and a Search for Quark Compositeness
in $\bbox{\bar{p}p}$ Collisions at $\bbox{\sqrt{s}}$ = 1.8 TeV
}
%
\author{                                                                      
B.~Abbott,$^{40}$                                                             
M.~Abolins,$^{37}$                                                            
V.~Abramov,$^{15}$                                                            
B.S.~Acharya,$^{8}$                                                           
I.~Adam,$^{39}$                                                               
D.L.~Adams,$^{48}$                                                            
M.~Adams,$^{24}$                                                              
S.~Ahn,$^{23}$                                                                
H.~Aihara,$^{17}$                                                             
G.A.~Alves,$^{2}$
% ...
\\                                                                            
\vskip 0.60cm                                                                 
\centerline{(D\O\ Collaboration)}                                             
\vskip 0.60cm                                                                 
}                                                                             
\address{                                                                     
\centerline{$^{1}$Universidad de Buenos Aires, Buenos Aires, Argentina}       
\centerline{$^{2}$LAFEX, Centro Brasileiro de Pesquisas F{\'\i}sicas,         
                  Rio de Janeiro, Brazil}                                     
\centerline{$^{3}$Universidade do Estado do Rio de Janeiro,                   
                  Rio de Janeiro, Brazil}                                     
\centerline{$^{4}$Universidad de los Andes, Bogot\'{a}, Colombia}             
\centerline{$^{5}$DAPNIA/Service de Physique des Particules, CEA, Saclay,     
                  France}                                                     
\centerline{$^{6}$Panjab University, Chandigarh, India}                       
\centerline{$^{7}$Delhi University, Delhi, India}                             
\centerline{$^{8}$Tata Institute of Fundamental Research, Mumbai, India}      
\centerline{$^{9}$Kyungsung University, Pusan, Korea}                         
\centerline{$^{10}$Seoul National University, Seoul, Korea}                   
\centerline{$^{11}$CINVESTAV, Mexico City, Mexico}                            
\centerline{$^{12}$Institute of Nuclear Physics, Krak\'ow, Poland}            
\centerline{$^{13}$Institute for Theoretical and Experimental Physics,        
                   Moscow, Russia}                                            
\centerline{$^{14}$Moscow State University, Moscow, Russia}                   
\centerline{$^{15}$Institute for High Energy Physics, Protvino, Russia}       
\centerline{$^{16}$University of Arizona, Tucson, Arizona 85721}              
\centerline{$^{17}$Lawrence Berkeley National Laboratory and University of    
                   California, Berkeley, California 94720}                    
\centerline{$^{18}$University of California, Davis, California 95616}         
\centerline{$^{19}$University of California, Irvine, California 92697}        
\centerline{$^{20}$University of California, Riverside, California 92521}     
\centerline{$^{21}$Florida State University, Tallahassee, Florida 32306}      
\centerline{$^{22}$University of Hawaii, Honolulu, Hawaii 96822}              
\centerline{$^{23}$Fermi National Accelerator Laboratory, Batavia,            
                   Illinois 60510}                                            
\centerline{$^{24}$University of Illinois at Chicago, Chicago,                
                   Illinois 60607}                                            
\centerline{$^{25}$Northern Illinois University, DeKalb, Illinois 60115}      
\centerline{$^{26}$Northwestern University, Evanston, Illinois 60208}         
\centerline{$^{27}$Indiana University, Bloomington, Indiana 47405}            
\centerline{$^{28}$University of Notre Dame, Notre Dame, Indiana 46556}       
\centerline{$^{29}$Purdue University, West Lafayette, Indiana 47907}          
\centerline{$^{30}$Iowa State University, Ames, Iowa 50011}                   
\centerline{$^{31}$University of Kansas, Lawrence, Kansas 66045}              
\centerline{$^{32}$Louisiana Tech University, Ruston, Louisiana 71272}        
\centerline{$^{33}$University of Maryland, College Park, Maryland 20742}      
\centerline{$^{34}$Boston University, Boston, Massachusetts 02215}            
\centerline{$^{35}$Northeastern University, Boston, Massachusetts 02115}      
\centerline{$^{36}$University of Michigan, Ann Arbor, Michigan 48109}         
\centerline{$^{37}$Michigan State University, East Lansing, Michigan 48824}   
\centerline{$^{38}$University of Nebraska, Lincoln, Nebraska 68588}           
\centerline{$^{39}$Columbia University, New York, New York 10027}             
\centerline{$^{40}$New York University, New York, New York 10003}             
\centerline{$^{41}$University of Rochester, Rochester, New York 14627}        
\centerline{$^{42}$State University of New York, Stony Brook,                 
                   New York 11794}                                            
\centerline{$^{43}$Brookhaven National Laboratory, Upton, New York 11973}     
\centerline{$^{44}$University of Oklahoma, Norman, Oklahoma 73019}            
\centerline{$^{45}$Brown University, Providence, Rhode Island 02912}          
\centerline{$^{46}$University of Texas, Arlington, Texas 76019}               
\centerline{$^{47}$Texas A\&M University, College Station, Texas 77843}       
\centerline{$^{48}$Rice University, Houston, Texas 77005}                     
}                                                                             

\date{16 July 1998}
\maketitle

\begin{abstract}
 Using the D\O\ detector at the 1.8 TeV \pbarp\ Fermilab Tevatron
 collider, we have measured the inclusive dijet mass spectrum in the
 central pseudorapidity region $\modetajet < 1.0$ for dijet masses
 greater than 200 \gevcc . We have also measured the ratio of spectra
 $\sigma(\modetajet < 0.5)/ \sigma(0.5 <\modetajet < 1.0)$. The order
 $\alpha_{\rm s}^{3}$ QCD predictions are in good agreement with the
 data and we rule out models of quark compositeness with a contact
 interaction scale $< 2.4$~TeV at the 95$\%$ confidence level.
\end{abstract}


 High transverse energy (\Et) jet production at a center of mass
 energy of 1.8 TeV probes the structure of the proton down to a
 distance scale of 10$^{-4}$ fm. A measurement of the dijet mass
 spectrum can be used to verify the predictions of quantum
 chromodynamics (QCD) for parton-parton scattering and to constrain
 the parton distribution functions (pdf) of the proton. Additionally,
 new physics such as quark compositeness~\cite{compositeness} would be
 revealed by an excess of events in the dijet mass (\jjmass ) spectrum
 at high masses with respect to the predictions of QCD. A previous
 analysis by the \mbox{CDF} collaboration of the inclusive jet cross section
 \cite{CDF} reported an excess of jet production at high \Et . More
 recent analyses of the dijet angular distribution by D\O
 ~\cite{d0_angular} and \mbox{CDF}~\cite{cdf_angular} have excluded at the
 95$\%$ confidence level models of quark compositeness in which the
 contact interaction scale is below 2 TeV. Most recently, an analysis
 of the inclusive jet cross section by D\O \cite{d0_inc} shows good
 agreement between the theory and data. This paper presents a new
 improved measurement by D\O\ of the inclusive dijet mass spectrum
 (uncertainty reduced by a factor of four at $\jjmass= 200$ GeV/$c^2$
 relative to the previous \mbox{CDF} measurement~\cite{cdf_dijet_mass}) and
 improved limits on the contact interaction scale.\\

 The outgoing partons from the parton-parton \mbox{scattering} process
 hadronize to form jets of particles. These jets were identified in
 the D\O\ detector~\cite{d0_detector} using uranium/liquid-argon
 calorimeters which cover a pseudorapidity range of $\modeta \leq 4.1$
 ($\eta = -{\rm ln}[{\rm tan}(\theta/2)]$, where $\theta$ is the polar
 angle relative to the proton beam direction).  The calorimeters have
 a jet transverse energy resolution of 10$\%$ (5$\%$) for \Et\ of 50
 (300) GeV.\\

 Events with at least one inelastic interaction during a beam crossing
 were identified using scintillator \mbox{hodoscopes}, and the primary
 event vertex was determined using tracks reconstructed in the central
 tracking system. Event selection occurred in two stages. First, a
 minimum transverse energy was required in a region ($\Delta
 \eta\times\Delta\phi=0.8\times 0.6$) of the calorimeter. Jet
 candidates were then reconstructed online with a cone algorithm of
 opening angle ${\cal{R}}=0.7$ in $\eta$--$\phi$ space ($\phi$ is the
 \mbox{azimuthal} angle), and the event was recorded if any jet \Et\
 \mbox{exceeded} a specified \mbox{threshold.}  During the
 \mbox{1994--95} run, the thresholds were 30, 50, 85, and 115 GeV,
 with integrated luminosities of 0.353$\pm$0.027, 4.69$\pm$0.37,
 54.7$\pm$3.4, and 91.9$\pm$5.6~\ipb , respectively. The luminosities
 of the 30 and 50 GeV triggers were determined by matching their dijet
 cross sections to that measured for the 85 GeV trigger.  This
 resulted in an additional uncertainty of 4.9$\%$ in the luminosities
 of the 30 and 50 GeV triggers.\\

 Jets were reconstructed offline using an iterative jet cone algorithm
 with ${\cal{R}}=0.7$~\cite{rsep}. Jet \Et\ is defined as the sum of
 the \Et\ in each cell within the cone. The jet was centered on the
 \Et -weighted pseudorapidity and azimuth of the jet.  The jet \Et\
 and direction were then \mbox{recalculated} until the cone direction
 was stable.  If \mbox{two} jets shared energy, they were combined or
 split based on the fraction of energy shared relative to the \Et\ of
 the \mbox{lower-\Et} jet. If the shared fraction exceeded $50\%$, the jets
 were combined and the direction recalculated. Otherwise, the jets
 were split and the energy in each of the shared cells was assigned to
 the nearest jet. The directions of the split jets were then
 recalculated~\cite{rsep}.\\

 A significant fraction of the data was taken at high
 \mbox{instantaneous} luminosity, which resulted in more than one
 \pbarp\ interaction in a beam crossing leading to an ambiguity in
 selecting the primary event vertex. After \mbox{event}
 reconstruction, the two vertices with the largest track multiplicity
 were retained. The quantity $\HT = \mid\!\!\Sigma \vec{E}_{T}^{\rm
 jet}\!\!\mid$ was calculated for both vertices, and the vertex with
 the smaller \HT\ was selected. The uncertainty on the mass spectrum
 due to the choice of vertex was 2$\%$. The vertex was required to be
 within 50 cm of the detector center. This cutoff was $90 \pm 1\%$
 efficient, independent of the dijet mass.\\

 Backgrounds from noise, cosmic rays, and accelerator losses were
 reduced to an insignificant level by applying jet quality
 criteria. For an event to be accepted, the two leading-\Et\ jets were
 required to satisfy these quality criteria. Contamination from
 backgrounds was $< 2\%$ based on Monte Carlo simulations and visual
 inspection of dijet events of high mass.  The overall jet selection
 efficiency for $\modeta \leq$ 1.0 was measured as a function of jet
 \Et , giving $97\pm 1\%$ at 250 GeV, and $94 \pm 1\%$ at 400 GeV.\\

 The transverse energy of each jet was corrected for the underlying
 event, additional interactions, calorimeter noise, the fraction of
 particle energy that showered hadronically outside of the cone, and
 for the hadronic \mbox{response~\cite{energy_scale}.} At $\eta=0$,
 the mean total jet energy correction was 16$\%$ (12$\%$) at 100~GeV
 (400~GeV); the correction uncertainty was less than 2.5$\%$ of the
 jet \Et .\\

 For each event that passed the criteria, the dijet mass, \jjmass ,
 was calculated, assuming that the jets are massless, using \mbox{$M^2
 = 2 \Etu{1} \Etu{2} [ \cosh( \Delta \eta ) - \cos ( \Delta \phi )
 ]$}.\\
 
 The steeply falling dijet mass spectrum is \mbox{distorted} by jet
 energy resolution (and to a negligible \mbox{extent} by $\eta$
 resolution).  The dijet mass resolution was calculated using the
 measured single-jet resolutions and the \mbox{{\sc
 pythia}~\cite{pythia}} Monte Carlo event generator.  This resolution
 depends on the \Et\ and $\eta$ distributions of the two leading \Et\
 jets in each event.  The observed mass spectrum was corrected with an
 ansatz function $F(M^{\prime}) = B {M^{\prime}}^{- \alpha} [ 1 -
 ({M^{\prime}}/{\sqrt{s}})]^{-\beta}$ convoluted with the mass
 resolutions, to obtain the smeared ansatz \mbox{$f(M) = \int^{\infty}_{0}
 F(M^{\prime})\rho(M^{\prime} - M, M^{\prime}) dM^{\prime}$} (where
 $\rho$ is the mass resolution), such that the number of events in any
 given mass bin $i$ is given by integrating $f$ over that bin. The
 data were then fitted using a binned maximum likelihood method and
 the {\sc minuit} \cite{minuit} package to determine the values of
 $B$, $\alpha$, and $\beta$. The unsmearing correction for each mass
 bin is then given by $C_{i} = { { \int F dM } / \int f dM }$ ($C_i =
 0.96$ (0.92) at 210 (900 GeV/$c^2$)).\\

 The dijet mass spectrum was calculated using {$$\kappa \equiv d^{3}
 \sigma/ d \jjmass d \eta_{1} d \eta_{2} = ( N_{i} C_{i} )/({\cal
 L}_{i} \epsilon \Delta \jjmass \Delta \eta_{1} \Delta \eta_{2}),$$
 where $N_i$ is the number of events in mass bin $i$, ${\cal L}_{i}$
 is the integrated luminosity for that bin, $\epsilon$ is the
 efficiency of the vertex selection and jet quality cuts, $\Delta
 \jjmass$ is the width of the mass bin, and $\Delta \eta_{1,2}$ is the
 width of the $\eta$ bin for jets 1 and 2. The spectrum was calculated
 for the pseudorapidity range $\modetajet < 1.0$ (where both jets are
 required to satisfy the $\eta$ requirement), in mass ranges starting
 at 200, 270, 350, and 550 \gevcc , corresponding to the software jet
 \Et\ thresholds of 30, 50, 85, and 115 GeV.\\

\newpage
 The cross section for the mass spectrum is plotted in
 Fig.~\ref{fig_1}, and given in Table~\ref{table_1}. The data are
 plotted at the mass-weighted average of the fit function for each bin
 (${ \int M F dM} / {\int F dM }$). The systematic uncertainties are
 dominated by the uncertainties in the jet energy \mbox{scale}, which
 are 7$\%$ (30$\%$) at 200 (1000) \gevcc . The \mbox{other}
 uncertainties are due to the luminosity \mbox{measurement} (6.1$\%$);
 luminosity matching at low mass (4.9$\%$); the \mbox{unsmearing}
 correction, 0.5$\%$ (3$\%$) at 200 (1000) \gevcc ; the vertex cut
 (1$\%$); and the jet selection cuts (1$\%$). The total systematic
 uncertainty is given by the sum of the individual uncertainties in
 quadrature. The bin-to-bin correlations of the uncertainties are
 shown in Fig.~\ref{fig_2}~\cite{correlations}.\\

\begin{figure}[hbtp]
\epsfxsize=5in
\centerline{\epsffile{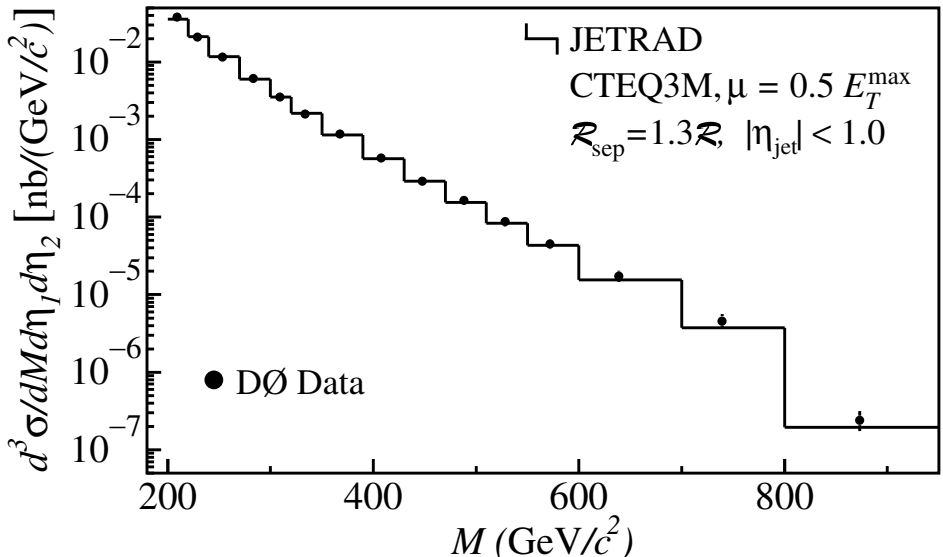}}
 \caption{$d^{3} \sigma / d \jjmass d \eta_{1} d \eta_{2}$ for
 $\modetajet<1.0$. The D\O\ data are shown by the solid circles, with
 error bars representing the $\pm$$1\sigma$ statistical and systematic
 uncertainties added in quadrature (in most cases smaller than the
 symbol). The histogram represents the {\sc jetrad} prediction.}
\label{fig_1}
\end{figure}

\begin{figure}[hbt]
\vspace{-2mm}
\epsfxsize=5in
\centerline{\epsffile{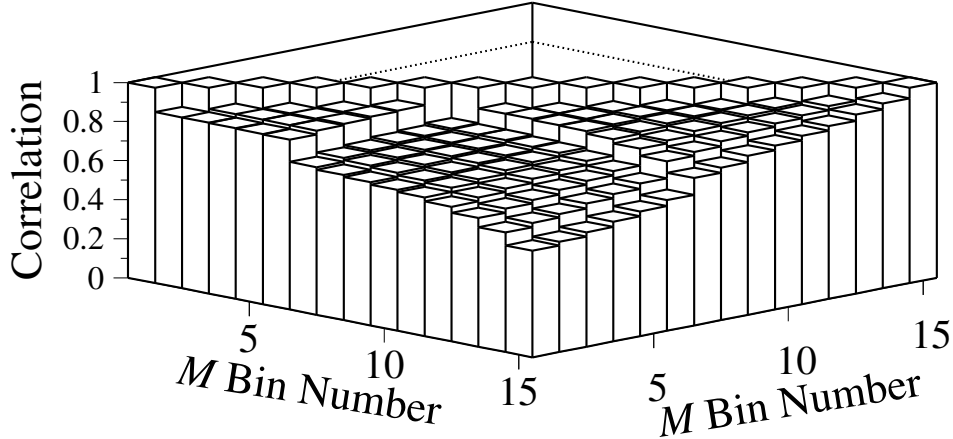}}
 \caption{The correlations between systematic uncertainties in bins of
 dijet mass (see Table~\ref{table_1}) for $\modetajet<1.0$. The
 correlations are calculated using the average systematic
 uncertainty. The discontinuities arise from the uncorrelated errors
 (adjacent to correlations of 1.0) and luminosity matching.}
\label{fig_2}
\end{figure}
 
\begin{table*}[htbp] 
\begin{center} 
\caption{Dijet cross section for $\modetajet < 1.0$, and the 
  ratio $\kappa(\modetajet < 0.5)/ \kappa(0.5 < \modetajet <
  1.0)$. High (low) systematic uncertainties are the sum in quadrature
  of the uncertainties from the $\pm1\sigma$ variations in the energy
  calibration, the unsmearing, the vertex corrections, luminosity
  matching, jet selection, and the uncertainty in the luminosity.  }
\label{table_1}
\begin{tabular}{rrrcr@{ $\pm$ }l@{$\times 10$}rddr@{ $\pm$ }l@{ $\pm$ }l}
\multicolumn{4}{c}{Mass Bin (\gevcc )}  &  
\multicolumn{5}{c}{$d^{3} \sigma / d \jjmass d \eta_{1} d \eta_{2}$ }  & 
\multicolumn{3}{c}{Ratio of Mass Spectra}\\
\cline{1-4}\cline{5-9}
\multicolumn{1}{c}{Bin} &
\multicolumn{1}{c}{Min.} & 
\multicolumn{1}{c}{Max.} &  
\multicolumn{1}{c}{Weighted} &
\multicolumn{3}{c}{$\pm$ Stat. Error} &  
\multicolumn{1}{c}{Syst. Low~\hspace*{2.25mm}} & 
\multicolumn{1}{c}{Syst. High} & 
\multicolumn{3}{c}{$\kappa(\modetajet < 0.5)/\kappa(0.5 <\modetajet < 1.0)$}\\
\multicolumn{1}{c}{}   & 
\multicolumn{1}{c}{}   & 
\multicolumn{1}{c}{}   & 
\multicolumn{1}{c}{Center} &
\multicolumn{3}{c}{(nb)}  & 
\multicolumn{1}{c}{($\%$)}  & 
\multicolumn{1}{c}{($\%$)}&
\multicolumn{3}{c}{($\pm$ stat. error $\pm$ syst. errror)} \\
\hline \hline
 1 &  200 &  220 &  209.1 &(3.78 & 0.12)&$^{-2}$& $-$11.4 &  +11.8 &  
\hspace*{10mm}0.613 &  0.039 &  0.037\\
 2 &  220 &  240 &  229.2 &(2.10 & 0.09)&$^{-2}$& $-$11.3 &  +11.6 &
  0.614 &  0.050 &  0.030\\
 3 &  240 &  270 &  253.3 &(1.16 & 0.06)&$^{-2}$& $-$11.5 &  +11.7 &
  0.570 &  0.051 &  0.029\\
 4 &  270 &  300 &  283.4 &(6.18 & 0.11)&$^{-3}$& $-$11.5 &  +12.0 &
  0.568 &  0.030 &  0.027\\
 5 &  300 &  320 &  309.3 &(3.55 & 0.11)&$^{-3}$& $-$11.5 &  +12.1 &
  0.610 &  0.034 &  0.050\\
 6 &  320 &  350 &  333.6 &(2.12 & 0.07)&$^{-3}$& $-$11.9 &  +12.3 &
  0.705 &  0.044 &  0.058\\
 7 &  350 &  390 &  367.6 &(1.18 & 0.01)&$^{-3}$& $-$11.1 &  +11.6 &
  0.672 &  0.020 &  0.032\\
 8 &  390 &  430 &  407.8 &(5.84 & 0.09)&$^{-4}$& $-$11.5 &  +12.2 &
  0.593 &  0.022 &  0.030\\
 9 &  430 &  470 &  447.9 &(2.89 & 0.06)&$^{-4}$& $-$11.9 &  +12.9 &
  0.708 &  0.036 &  0.037\\
10 &  470 &  510 &  488.0 &(1.64 & 0.05)&$^{-4}$& $-$12.4 &  +13.5 &
  0.690 &  0.046 &  0.036\\
11 &  510 &  550 &  528.0 &(8.74 & 0.34)&$^{-5}$& $-$12.8 &  +14.3 &
  0.620 &  0.058 &  0.033\\
12 &  550 &  600 &  572.0 &(4.49 & 0.17)&$^{-5}$& $-$13.5 &  +15.3 &
  0.634 &  0.065 &  0.033\\
13 &  600 &  700 &  638.9 &(1.73 & 0.07)&$^{-5}$& $-$14.9 &  +17.2 &
  0.647 &  0.074 &  0.034\\
14 &  700 &  800 &  739.2 &(4.58 & 0.38)&$^{-6}$& $-$17.6 &  +20.8 &
  0.608 &  0.141 &  0.035\\
15 &  800 & 1400 &  873.2 &(2.39 & 0.35)&$^{-7}$& $-$23.2 &  +28.9 &
  0.705 &  0.246 &  0.046\\
\end{tabular}
\end{center}
\end{table*}

 The histogram in Fig.~\ref{fig_1} is a prediction for the inclusive
 dijet mass spectrum from the next-to-leading \mbox{(NLO)} parton
 level event generator {\sc jetrad}~\cite{jetrad}.  The NLO
 calculation requires specification of the renormalization and
 factorization scales ($\mu = 0.5 \Etmax$ where \Etmax\ is the maximum
 jet \Et\ in the generated event), pdf (CTEQ3M \cite{cteq}), and
 parton clustering algorithm.  Two partons are combined if they are
 within ${\cal{R}}_{\rm sep}=1.3{\cal{R}}$, as \mbox{motivated} by the
 separation of jets in the data~\cite{rsep}.  Choosing an alternative
 pdf (CTEQ4M~\cite{cteq4m}, \mbox{CTEQ4HJ~\cite{cteq4m}}, or
 MRS(A$^\prime$)~\cite{mrsap}) \mbox{alters} the prediction by as much
 as $25\%$, and varying $\mu$ in the range $0.25\Etmax$ to $2\Etmax$
 alters the normalization by up to $30\%$ with some \jjmass\
 dependence. The CTEQ3M and MRS(A$^\prime$) pdf's are fits to collider
 and fixed target data sets published before 1994. CTEQ4M updates
 these fits using data published before 1996, and CTEQ4HJ
 \mbox{adjusts} the gluon distributions to fit the \mbox{CDF}
 inclusive jet cross section measurement~\cite{CDF}.
 Figure~\ref{fig_3} shows the ratio $({\rm Data} - {\rm Theory})/{\rm
 Theory}$ for the {\sc jetrad} prediction using CTEQ3M with $\mu =
 0.5\Etmax$. Given the experimental and theoretical uncertainties, the
 predictions can be regarded as in good agreement with the data. The
 data are also in agreement within the given uncertainties with the
 cross section measured by \mbox{CDF}~\cite{cdf_dijet_mass}.\\

\begin{figure}[htb]
\epsfxsize=5in
\centerline{\epsffile{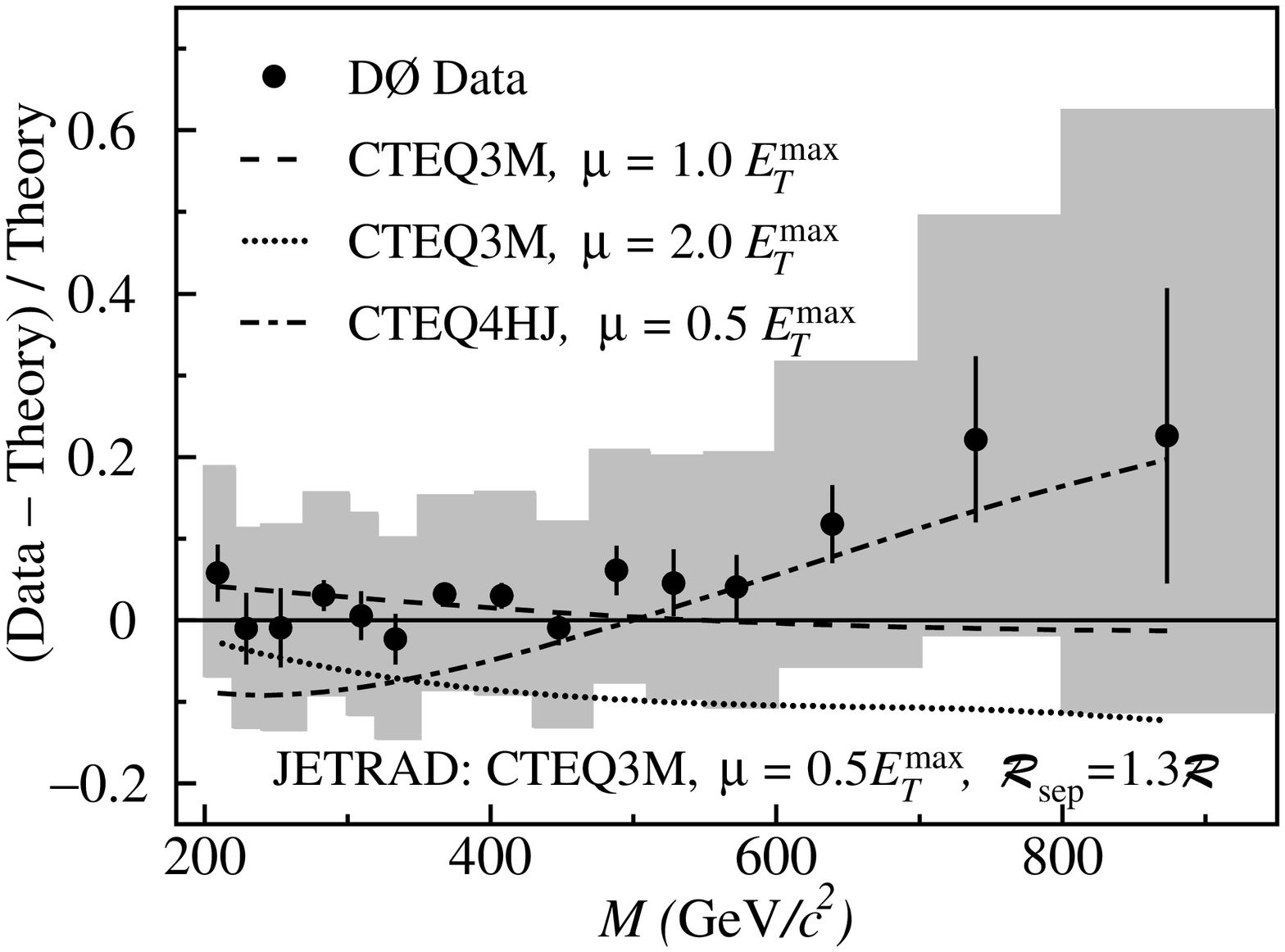}}
\caption{The difference between the data and the  prediction ({\sc jetrad})
 divided by the prediction for $\modetajet < 1.0$. The solid circles
 represent the comparison to the calculation using CTEQ3M with $\mu =
 0.5 \Etmax$. The shaded region represents the $\pm$$1\sigma$
 systematic uncertainties. The effects of changing the renormalization
 scale and choosing a different pdf are also shown (each curve shows
 the difference between the alternative prediction and the standard
 prediction).}
\label{fig_3}
\end{figure}

 In Table~\ref{table_2}, we show the \chisq\ resulting from a fit of
 theory to our data, using the full correlation matrix \mbox{between}
 different mass bins.  The choice of pdf and renormalization scale is
 varied; all choices give reasonable probability.\\

\begin{table}[htbp]
 \caption{\chisq\ values calculated for various theoretical
 predictions for the dijet mass spectrum with $\modetajet<1.0$, and
 for the ratio of cross sections (15 degrees of freedom).}
 \label{table_2} \begin{center} \begin{tabular}{ccdddd} pdf & D &
\multicolumn{2}{c}{~Mass Spectrum} &  
\multicolumn{2}{c}{Ratio} \\
      & where $\mu =D \Etmax$  &   
\multicolumn{1}{c}{$\chisq$} & 
\multicolumn{1}{c}{Prob.} & 
\multicolumn{1}{c}{$\chisq$} & 
\multicolumn{1}{c}{Prob.}       \\
\hline\hline
   CTEQ3M         & $0.25$ &  12.2 & 0.66 & 40.5 & 0.00\\ 
   CTEQ3M         & $0.50$ &   5.0 & 0.99 & 15.9 & 0.39\\ 
   CTEQ3M         & $0.75$ &   5.3 & 0.99 & 14.7 & 0.48\\ 
   CTEQ3M         & $1.00$ &   5.4 & 0.99 & 14.3 & 0.51\\ 
   CTEQ3M         & $2.00$ &   4.2 & 1.00 & 13.7 & 0.55\\ 
   CTEQ4M         & $0.50$ &   4.9 & 0.99 & 15.7 & 0.40\\ 
   CTEQ4HJ        & $0.50$ &   5.0 & 0.99 & 16.0 & 0.38\\ 
   MRS(A$^\prime$)& $0.50$ &   6.3 & 0.97 & 16.3 & 0.36\\ 
  \end{tabular}
 \end{center}
\end{table}

 The ratio \mbox{$\kappa(\modetajet<0.5)/\kappa(0.5<\modetajet<1.0)$},
 given in Fig.~\ref{fig_4} and Table~\ref{table_1}, exploits the high
 correlation between uncertainties in the measurement of the dijet
 mass spectrum.  The resulting cancelation of uncertainties leads to a
 systematic error of less than 8$\%$ for \mbox{all} \jjmass . The
 uncertainty in the theoretical prediction of this ratio is less than
 $3\%$ due to the choice of pdf, and $6\%$ from the choice of
 renormalization and factorization \mbox{scale} (excluding $\mu = 0.25
 \Etmax$). The \chisq\ values are shown in Table~\ref{table_2}.  The
 predictions are in good agreement with the data, \mbox{except} for
 $\mu = 0.25 \Etmax$ which is excluded by the data (\chisq\ = 40.5, a
 probability of 0.04$\%$).\\
 
 The ratio of the mass spectra can be used to place limits on quark
 compositeness. A mass scale $\Lambda$ characterizes both the strength
 of the quark-substructure coupling and the physical size of the
 composite state. Limits are set assuming that
 \mbox{$\Lambda\gg\sqrt{\hat{s}}$} (where $\sqrt{\hat{s}}$ is the
 center of mass energy of the colliding partons), such that
 \mbox{quarks} appear to be point-like. Hence, the substructure
 coupling can be approximated by a four-Fermi \mbox{contact}
 interaction giving rise to an effective
 Lagrangian~\cite{compositeness} \mbox{${\cal L} =
 A(2\pi/\Lambda^{2})(\bar{q}_{L}\gamma^{\mu}q_{L})
 (\bar{q}_{L}\gamma_{\mu}q_{L})$}, where $A=\pm1$, and $q_L$
 represents left-handed quarks. Limits are presented for the case
 where all quarks are composite, showing both constructive
 interference ($\Lambda^{-}$ for $A=-1$) and destructive interference
 ($\Lambda^{+}$ for $A=+1$). Currently there are no NLO compositeness
 calculations available; therefore, the {\sc pythia} event generator
 is used to simulate the effect of compositeness. The ratio of these
 LO predictions with compositeness, to the LO with no compositeness,
 is used to scale the {\sc jetrad} NLO prediction, as shown in
 Fig.~\ref{fig_4}.\\

\begin{figure}[htb]
\epsfxsize=5in
\centerline{\epsffile{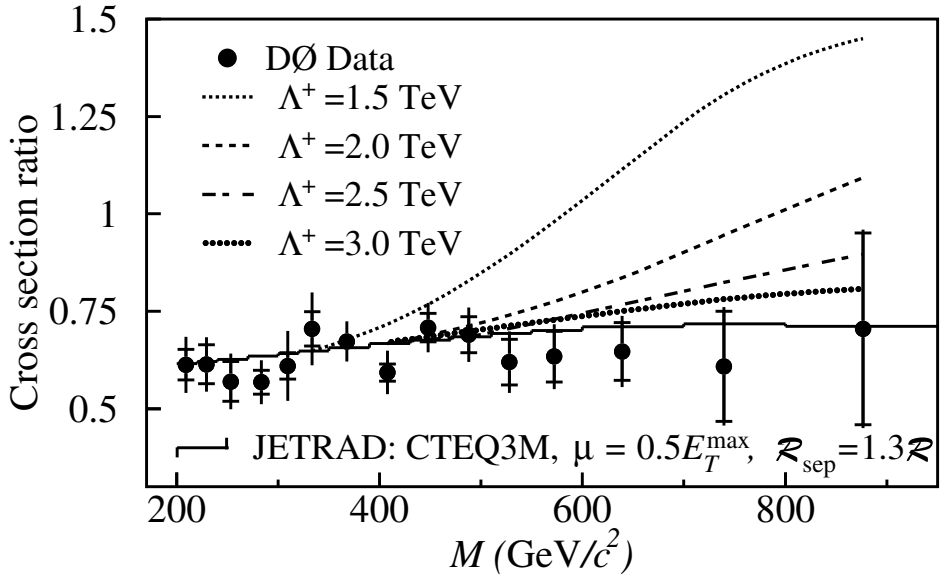}}
 \caption{The ratio of cross sections for $\modetajet<0.5$ and
 $0.5<\modetajet <1.0$ for data (solid circles) and theory (various
 lines).  The error bars show the statistical and systematic
 uncertainties added in quadrature, and the crossbar shows the size of
 the statistical error.  }
\label{fig_4}
\end{figure}

 We employ a Bayesian technique \cite{Bayes} to obtain from our data a
 limit on the scale of compositeness. Motivated by the form of the
 Lagrangian, a uniform prior is assumed in $\xi=1/\Lambda^2$, and a
 Gaussian likelihood function $P \propto e^{-\chi^{2}/2}$ is used.
 The 95$\%$ confidence limit in $\Lambda$ is determined by requiring
 that $\int_{0}^{\xi} P(\xi')d\xi'=0.95$. Since the ratio at NLO is
 sensitive to the choice of $\mu$ and pdf, each possible choice is
 treated as a different theory. The most conservative lower limits on
 the mass scale at the 95$\%$ confidence level are found to be
 $\Lambda^{+}>2.7$~TeV and $\Lambda^{-}>2.4$~TeV. These limits are
 incompatible with the suggestion of a compositeness scale $\Lambda$
 in the 1.5 to 1.8 TeV range found from earlier
 measurements~\cite{CDF} of the high \Et\ jet inclusive cross section.\\

 In conclusion, we have measured the cross section for the inclusive
 dijet mass spectrum for $\modetajet<1.0$ with $\jjmass>200$~\gevcc ,
 and the ratio of cross sections for $\modetajet<0.5$ and
 $0.5<\modetajet<1.0$, as a function of dijet mass.  The data
 distributions are in good agreement with NLO QCD predictions. Models
 of quark compositeness with a contact interaction scale of less than
 2.4~TeV are excluded at the 95$\%$ confidence level.\\

%
 We thank the staffs at Fermilab and collaborating institutions for
 their contributions to this work, and acknowledge support from the
 Department of Energy and National Science Foundation (U.S.A.),
 Commissariat \` a L'Energie Atomique (France), State Committee for
 Science and Technology and Ministry for Atomic Energy (Russia), CAPES
 and CNPq (Brazil), Departments of Atomic Energy and Science and
 Education (India), Colciencias (Colombia), CONACyT (Mexico), Ministry
 of Education and KOSEF (Korea), and CONICET and UBACyT (Argentina).
 We thank W.T. Giele, E.W.N. Glover, and D.A. Kosower for help with
 {\sc jetrad}.

\end{document}